\documentclass[10pt]{article}
\usepackage{graphicx}
\usepackage{amsmath}
\usepackage{amssymb}
\usepackage{caption2}
\begin{document}
\bibliographystyle{prsty}
\begin{center}
{\large {\bf \sc{  Analysis of  the $P_{cs}(4459)$ as the hidden-charm pentaquark state  with QCD sum rules }}} \\[2mm]
Zhi-Gang Wang \footnote{E-mail: zgwang@aliyun.com.  }     \\
 Department of Physics, North China Electric Power University, Baoding 071003, P. R. China
\end{center}

\begin{abstract}
In this article, we study the scalar-diquark-scalar-diquark-antiquark type $udsc\bar{c}$ pentaquark state with the QCD sum rules, the predicted mass $M_P=4.47\pm0.11\,\rm{GeV}$ is in excellent agreement with the experimental data $ 4458.8 \pm 2.9 {}^{+4.7}_{-1.1} \mbox{ MeV}$ from the  LHCb collaboration, and supports assigning the $P_{cs}(4459)$   to be the hidden-charm  pentaquark state
with the spin-parity $J^P={\frac{1}{2}}^-$. We take into account the flavor $SU(3)$ mass-breaking effect, and estimate the mass spectrum of the diquark-diquark-antiquark type $udsc\bar{c}$ pentaquark states with the strangeness $S=-1$.
\end{abstract}



\section{Introduction}
In 2015,  the  LHCb collaboration  observed  two pentaquark candidates $P_c(4380)$ and $P_c(4450)$ in the $J/\psi p$ mass spectrum     with the statistical significances larger  than $9\sigma$ in the $\Lambda_b^0\to J/\psi K^- p$ decays \cite{LHCb-4380}.
In 2019, the LHCb collaboration  observed a  narrow pentaquark candidate $P_c(4312)$  with a statistical significance of  $7.3\sigma$ in the $\Lambda_b^0\to J/\psi K^- p$ decays,  and proved that the $P_c(4450)$ consists  of two narrow overlapping peaks $P_c(4440)$ and $P_c(4457)$
  with  the statistical significance of  $5.4\sigma$ \cite{LHCb-Pc4312}.
  The Breit-Wigner  masses and widths are
\begin{flalign}
 &P_c(4312) : M = 4311.9\pm0.7^{+6.8}_{-0.6} \mbox{ MeV}\, , \, \Gamma = 9.8\pm2.7^{+ 3.7}_{- 4.5} \mbox{ MeV} \, , \nonumber \\
 & P_c(4440) : M = 4440.3\pm1.3^{+4.1}_{-4.7} \mbox{ MeV}\, , \, \Gamma = 20.6\pm4.9_{-10.1}^{+ 8.7} \mbox{ MeV} \, , \nonumber \\
 &P_c(4457) : M = 4457.3\pm0.6^{+4.1}_{-1.7} \mbox{ MeV} \, ,\, \Gamma = 6.4\pm2.0_{- 1.9}^{+ 5.7} \mbox{ MeV} \,   .
\end{flalign}
 Very recently, the LHCb collaboration reported an evidence of a hidden-charm pentaquark candidate $P_{cs}(4459)$ with the strangeness $S=-1$ in the $J/\psi \Lambda$ invariant mass spectrum with  a statistical significance of  $3.1\sigma$ in the $\Xi_b^- \to J/\psi K^- \Lambda$ decays  \cite{LHCb-Pcs4459,LHCb-Pcs4459-2012},
the Breit-Wigner mass and width are
\begin{flalign}
 &P_{cs}(4459) : M = 4458.8 \pm 2.9 {}^{+4.7}_{-1.1} \mbox{ MeV}\, , \, \Gamma = 17.3 \pm 6.5 {}^{+8.0}_{-5.7} \mbox{ MeV} \, ,
\end{flalign}
but the spin and parity have not been determined yet.
 There have been  several  possible  assignments for  those hidden-charm  $P_c$ states, such as the diquark-diquark-antiquark type pentaquark states \cite{di-di-anti-penta-1,di-di-anti-penta-2,di-di-anti-penta-3,di-di-anti-penta-4,di-di-anti-penta-5,Wang1508-EPJC,WangHuang-EPJC-1508-12,
 WangZG-EPJC-1509-12,WangZG-NPB-1512-32,WangZhang-APPB,WZG-penta-IJMPA,Pc4312-penta-1,Pc4312-penta-2,Pc4312-penta-3},  the diquark-triquark  type  pentaquark states \cite{di-tri-penta-1,di-tri-penta-2,di-tri-penta-3},
 the  molecule-like  pentaquark  states \cite{mole-penta-before-Wu-PRL,mole-penta-before-Wu-PRC,mole-penta-before-Zhu,mole-penta-1,mole-penta-2,mole-penta-3,mole-penta-4,mole-penta-5,mole-penta-6,mole-penta-7,
 mole-penta-8,mole-penta-9,mole-penta-10,mole-penta-11,mole-penta-12,mole-penta-13,Pc4312-mole-penta-1,Pc4312-mole-penta-2,Pc4312-mole-penta-3,Pc4312-mole-penta-4,Pc4312-mole-penta-5,Pc4312-mole-penta-6,Pc4312-mole-penta-7,Pc4312-mole-penta-8,
Pc4312-mole-penta-9,Pc4312-mole-penta-10,Pcs4459-mole-penta-1,Pcs4459-mole-penta-2,Pcs4459-mole-penta-3}, the hadro-charmonium states \cite{Pc4312-hadrocharmonium-1,Pc4312-hadrocharmonium-2,Pc4312-hadrocharmonium-3},
 the re-scattering effects \cite{rescattering-penta-1,rescattering-penta-2}, etc.

 In Refs.\cite{Wang1508-EPJC,WangHuang-EPJC-1508-12,WangZG-EPJC-1509-12,WangZG-NPB-1512-32,WangZhang-APPB}, we construct the (light)diquark-(heavy)diquark-antiquark type pentaquark currents to study the spin-parity $J^P={\frac{1}{2}}^\pm$, ${\frac{3}{2}}^\pm$, ${\frac{5}{2}}^\pm$   hidden-charm pentaquark states with the strangeness  $S=0,\,-1,\,-2,\,-3$ via the QCD sum rules in an systematic way   by taking into account the contributions of the vacuum condensates up to  dimension 10, and explore the possible assignments of the $P_c(4380)$ and $P_c(4450)$. Now we confront the predicted masses of the pentaquark states with the strangeness $S=-1$ to the LHCb data \cite{LHCb-Pcs4459,LHCb-Pcs4459-2012}.

 In the case of the axialvector-diquark-scalar-diquark-antiquark type pentaquark states with the strangeness  $S=-1$, the predicted masses $M=4.51\pm0.11\,\rm{GeV}$ and $4.51\pm0.12\,\rm{GeV}$ for the pentaquark states with the spin-parity $J^P={\frac{1}{2}}^-$ and ${\frac{3}{2}}^-$ respectively are compatible with the LHCb data $4458.8 \pm 2.9 {}^{+4.7}_{-1.1} \mbox{ MeV}$ within uncertainties \cite{WangZG-EPJC-1509-12,WangZG-NPB-1512-32}.

   In the case of the axialvector-diquark-axialvector-diquark-antiquark type pentaquark states with the strangeness  $S=-1$, the predicted masses $M=4.47\pm0.15\,\rm{GeV}$, $4.51\pm 0.12\,\rm{GeV}$  and $4.52\pm0.12\,\rm{GeV}$ for the pentaquark states with the spin-parity $J^P={\frac{1}{2}}^-$, ${\frac{3}{2}}^-$ and ${\frac{3}{2}}^-$ respectively are also compatible with the LHCb data $4458.8 \pm 2.9 {}^{+4.7}_{-1.1} \mbox{ MeV}$ within uncertainties \cite{WangZG-EPJC-1509-12,WangZG-NPB-1512-32}.

  In the case of the scalar-diquark-axialvector-diquark-antiquark type pentaquark states with the strangeness  $S=-1$, the predicted mass $M=4.49\pm0.04\,\rm{GeV}$ for the pentaquark state with the spin-parity $J^P={\frac{3}{2}}^-$ is also compatible with the LHCb data $4458.8 \pm 2.9 {}^{+4.7}_{-1.1} \mbox{ MeV}$ within uncertainties \cite{WangZhang-APPB}.

  There are six possible pentaquark configurations with the spin-parity $J^P={\frac{1}{2}}^-$ or ${\frac{3}{2}}^-$ for the $P_{cs}(4459)$ in the old calculations, where
 we took into account the contributions of the vacuum condensates up to dimension  $10$, and used the energy scale formula $\mu=\sqrt{M_{P}^2-(2{\mathbb{M}}_c)^2}$  with the old value ${\mathbb{M}}_c=1.80\,\rm{GeV}$  to choose the optimal energy scales of the QCD spectral densities \cite{WangZG-EPJC-1509-12,WangZG-NPB-1512-32,WangZhang-APPB}.

After the discovery of the $P_c(4312)$,  we updated the old analysis by carrying out the operator product expansion up to the   vacuum condensates of $13$ in a consistent way and choosing the new value of the effective $c$-quark mass ${\mathbb{M}}_c=1.82\,\rm{GeV}$ \cite{WZG-tetraquark-Mc},  and  restudied the ground state mass spectrum of the diquark-diquark-antiquark type $uudc\bar{c}$ pentaquark states with the QCD sum rules  \cite{WZG-penta-IJMPA}.
The  predicted pentaquark  masses are improved slightly or remarkably  when the vacuum condensates of dimensions $11$ and $13$ are included, the lowest pentaquark state is $[ud][uc]\bar{c}$ ($0$, $0$, $0$, $\frac{1}{2}$)  with the spin-parity $J^P={\frac{1}{2}}^-$, which happens to  coincide with  the $P_c(4312)$.
Furthermore, the predicted pentaquark masses  support  assigning the $P_c(4440)$ to be the hidden-charm pentaquark state with  $J^{P}={\frac{1}{2}}^-$, ${\frac{3}{2}}^-$ or ${\frac{5}{2}}^-$ and assigning the $P_c(4457)$ to be the hidden-charm pentaquark state with $J^{P}={\frac{1}{2}}^-$ or ${\frac{3}{2}}^-$ \cite{WZG-penta-IJMPA}.

Now we take into account the flavor $SU(3)$ mass-breaking effects. For the traditional charmed baryon states, the  flavor $SU(3)$ mass-breaking effects due to the $s$-quark are $\Delta\approx 190\,\rm{MeV}$ and $120\,\rm{MeV}$ for the flavor antitriplets and sextets respectively from the Particle Data Group \cite{PDG}. We can take the average mass-breaking effects as  $\Delta=m_s=150\,\rm{MeV}$ in the present work, and make a crude estimation  for the masses of the pentaquark states with the strangeness  $S=-1$, which are presented in Table \ref{mass-1508-et al-Pcs}. From the table, we can see that the estimated  mass $4.46\pm0.11\,\rm{GeV}$ for the $[ud][sc]\bar{c}$ ($0$, $0$, $0$, $\frac{1}{2}$) pentaquark state is in excellent agreement with the experimental data $ 4458.8 \pm 2.9 {}^{+4.7}_{-1.1} \mbox{ MeV}$ from the  LHCb collaboration  \cite{LHCb-Pcs4459,LHCb-Pcs4459-2012}.

In this article, we extend our previous works \cite{Wang1508-EPJC,WangHuang-EPJC-1508-12,
 WangZG-EPJC-1509-12,WangZG-NPB-1512-32,WangZhang-APPB,WZG-penta-IJMPA} to study the ground state mass  of the diquark-diquark-antiquark type $udsc\bar{c}$ pentaquark state with the QCD sum rules, and explore the possible assignment of the $P_{cs}(4459)$ as the $[ud][sc]\bar{c}$ ($0$, $0$, $0$, $\frac{1}{2}$) pentaquark state with the spin-parity $J^P={\frac{1}{2}}^-$.

\begin{table}
\begin{center}
\begin{tabular}{|c|c|c|c|c|c|c|c|c|}\hline\hline
$[qq^\prime][q^{\prime\prime}c]\bar{c}$ ($S_L$, $S_H$, $J_{LH}$, $J$) &New analysis \cite{WZG-penta-IJMPA}    &$u\,{\rm or}\,d \to s$  &Assignments        \\ \hline

$[ud][uc]\bar{c}$ ($0$, $0$, $0$, $\frac{1}{2}$)                      &$4.31\pm0.11$   &$4.46\pm0.11$           &$?\,P_{cs}(4459)$         \\

$[ud][uc]\bar{c}$ ($0$, $1$, $1$, $\frac{1}{2}$)                      &$4.45\pm0.11$   &$4.60\pm0.11$           &     \\

$[uu][dc]\bar{c}+2[ud][uc]\bar{c}$ ($1$, $0$, $1$, $\frac{1}{2}$)     &$4.46\pm0.11$   &$4.61\pm0.11$           &      \\

$[uu][dc]\bar{c}+2[ud][uc]\bar{c}$ ($1$, $1$, $0$, $\frac{1}{2}$)     &$4.34\pm0.14$   &$4.49\pm0.14$           &      \\

$[ud][uc]\bar{c}$ ($0$, $1$, $1$, $\frac{3}{2}$)                      &$4.39\pm0.11$   &$4.54\pm0.11$           &       \\

$[uu][dc]\bar{c}+2[ud][uc]\bar{c}$ ($1$, $0$, $1$, $\frac{3}{2}$)     &$4.47\pm0.11$   &$4.62\pm0.11$           &    \\

$[uu][dc]\bar{c}+2[ud][uc]\bar{c}$ ($1$, $1$, $2$, $\frac{3}{2}$)     &$4.61\pm0.11$   &$4.76\pm0.11$           &                      \\

$[uu][dc]\bar{c}+2[ud][uc]\bar{c}$ ($1$, $1$, $2$, $\frac{3}{2}$)     &$4.52\pm0.11$   &$4.67\pm0.11$           &                     \\

$[uu][dc]\bar{c}+2[ud][uc]\bar{c}$ ($1$, $1$, $2$, $\frac{5}{2}$)     &$4.39\pm0.11$   &$4.54\pm0.11$           &      \\

$[ud][uc]\bar{c}$ ($0$, $1$, $1$, $\frac{5}{2}$)                      &$4.39\pm0.11$   &$4.54\pm0.11$           &     \\ \hline\hline
\end{tabular}
\end{center}
\caption{ The masses (in unit of GeV) of the pentaquark states with the strangeness $S=-1$, where the $S_L$ and $S_H$ denote the spins of the light diquarks and heavy diquarks respectively, $\vec{J}_{LH}=\vec{S}_L+\vec{S}_H$, $\vec{J}=\vec{J}_{LH}+\vec{J}_{\bar{c}}$, the $\vec{J}_{\bar{c}}$ is the angular momentum of the $\bar{c}$-quark.  }\label{mass-1508-et al-Pcs}
\end{table}

 The article is arranged as follows:   we derive the QCD sum rules for the mass and pole residue of  the
$[ud][sc]\bar{c}$ ($0$, $0$, $0$, $\frac{1}{2}$) pentaquark state in Sect.2;  in Sect.3, we present the numerical results and discussions; and Sect.4 is reserved for our
conclusion.

\section{QCD sum rules for  the  $udsc\bar{c}$ pentaquark state}
Firstly, let us write down  the two-point correlation function $\Pi(p)$ in the QCD sum rules,
\begin{eqnarray}\label{CF-Pi-Pi-Pi}
\Pi(p)&=&i\int d^4x e^{ip \cdot x} \langle0|T\left\{J(x)\bar{J}(0)\right\}|0\rangle \, ,
\end{eqnarray}
where
\begin{eqnarray}
 J(x)&=&\varepsilon^{ila} \varepsilon^{ijk}\varepsilon^{lmn}  u^T_j(x) C\gamma_5 d_k(x)\,s^T_m(x) C\gamma_5 c_n(x)\,  C\bar{c}^{T}_{a}(x) \, ,
\end{eqnarray}
where the $i$, $j$, $k$, $l$, $m$, $n$ and $a$ are color indices, the $C$ is the charge conjugation matrix \cite{Wang1508-EPJC,WangHuang-EPJC-1508-12,WangZG-EPJC-1509-12,WangZG-NPB-1512-32}. We choose the scalar-diquark-scalar-diquark-antiquark type current to interpolate the  $[ud][sc]\bar{c}$ ($0$, $0$, $0$, $\frac{1}{2}$) pentaquark state, see Table \ref{mass-1508-et al-Pcs}.
The current $J(x)$ couples potentially to the hidden-charm pentaquark state with the negative parity,
\begin{eqnarray}\label{Coupling-N}
\langle 0| J (0)|P^{-}(p)\rangle &=&\lambda_{-} U^{-}(p,s) \, ,
\end{eqnarray}
furthermore, it also couples potentially to the hidden-charm pentaquark state with the positive  parity,
\begin{eqnarray}\label{Coupling-P}
\langle 0| J (0)|P^{+}(p)\rangle &=&\lambda_{+} i\gamma_5 U^{+}(p,s) \, ,
\end{eqnarray}
as   multiplying $i \gamma_{5}$ to the current  $J(x)$ changes its parity \cite{Wang1508-EPJC,WangHuang-EPJC-1508-12,WangZG-EPJC-1509-12,WangZG-NPB-1512-32,WZG-penta-IJMPA,Chung82,Bagan93,Oka96,
WangHbaryon-1,WangHbaryon-2,WangHbaryon-3,WangHbaryon-4,WangHbaryon-5,WangHbaryon-6,Wang-cc-baryon-penta}, where the $U^\pm(p,s)$ are the Dirac spinors.

 At the hadron  side, we insert  a complete set  of intermediate pentaquark states with the same quantum numbers as the current operators  $J(x)$  and $i\gamma_5 J(x)$ into the correlation function
$\Pi(p)$ to obtain the hadronic representation
\cite{SVZ79-1,SVZ79-2,PRT85}, and  isolate the  lowest negative-parity and positive-parity hidden-charm pentaquark states, and obtain the
 results:
\begin{eqnarray}\label{CF-Hadron}
\Pi(p) & = & \lambda_{-}^2  {\!\not\!{p}+ M_{-} \over M_{-}^{2}-p^{2}  }+  \lambda_{+}^2  {\!\not\!{p}- M_{+} \over M_{+}^{2}-p^{2}  } +\cdots  \, ,\nonumber\\
&=&\Pi^1(p^2)\!\not\!{p}+\Pi^0(p^2)\, .
 \end{eqnarray}

 We obtain the spectral densities at the hadron  side through  dispersion relation,
\begin{eqnarray}
\frac{{\rm Im}\Pi^1(s)}{\pi}&=& \lambda_{-}^2 \delta\left(s-M_{-}^2\right)+\lambda_{+}^2 \delta\left(s-M_{+}^2\right) =\, \rho^1_{H}(s) \, , \\
\frac{{\rm Im}\Pi^0(s)}{\pi}&=&M_{-}\lambda_{-}^2 \delta\left(s-M_{-}^2\right)-M_{+}\lambda_{+}^2 \delta\left(s-M_{+}^2\right)
=\rho^0_{H}(s) \, ,
\end{eqnarray}
where we add the subscript $H$ to denote  the hadron side,
then we introduce the  weight functions $\sqrt{s}\exp\left(-\frac{s}{T^2}\right)$ and $\exp\left(-\frac{s}{T^2}\right)$ to obtain the QCD sum rules
at the hadron side,
\begin{eqnarray}
\int_{4m_c^2}^{s_0}ds \left[\sqrt{s}\,\rho^1_{H}(s)+\rho^0_{H}(s)\right]\exp\left( -\frac{s}{T^2}\right)
&=&2M_{-}\lambda_{-}^2\exp\left( -\frac{M_{-}^2}{T^2}\right) \, ,
\end{eqnarray}
where the $s_0$ is the continuum threshold parameter, and the $T^2$ is the Borel parameter.
We take the combinations  $\sqrt{s}\,\rho^1_{H}(s)\pm\rho^0_{H}(s)$ to separate the  contributions  of the pentaquark  states with the negative parity and positive parity unambiguously.

At the QCD side,  we contract the $u$, $d$, $s$ and $c$ quark fields in the correlation function $\Pi(p)$  with Wick theorem,
\begin{eqnarray}\label{QCD-Pi}
\Pi(p)&=&-i\,\varepsilon^{ila}\varepsilon^{ijk}\varepsilon^{lmn}\varepsilon^{i^{\prime}l^{\prime}a^{\prime}}\varepsilon^{i^{\prime}j^{\prime}k^{\prime}}
\varepsilon^{l^{\prime}m^{\prime}n^{\prime}}\int d^4x e^{ip\cdot x} \nonumber\\
&&   {\rm Tr}\Big[\gamma_5 D_{kk^\prime}(x) \gamma_5 C U^{T}_{jj^\prime}(x)C\Big] \,{\rm Tr}\Big[\gamma_5 C_{nn^\prime}(x) \gamma_5 C S^{T}_{mm^\prime}(x)C\Big] C C_{a^{\prime}a}^T(-x)C   \, ,
\end{eqnarray}
where
the $U_{ij}(x)$, $D_{ij}(x)$, $S_{ij}$ and $C_{ij}(x)$ are the full $u$, $d$, $s$ and $c$ quark propagators, respectively,
 \begin{eqnarray}
U/D_{ij}(x)&=& \frac{i\delta_{ij}\!\not\!{x}}{ 2\pi^2x^4}-\frac{\delta_{ij}\langle
\bar{q}q\rangle}{12} -\frac{\delta_{ij}x^2\langle \bar{q}g_s\sigma Gq\rangle}{192} -\frac{ig_sG^{a}_{\alpha\beta}t^a_{ij}(\!\not\!{x}
\sigma^{\alpha\beta}+\sigma^{\alpha\beta} \!\not\!{x})}{32\pi^2x^2} -\frac{\delta_{ij}x^4\langle \bar{q}q \rangle\langle g_s^2 GG\rangle}{27648} \nonumber\\
&&  -\frac{1}{8}\langle\bar{q}_j\sigma^{\mu\nu}q_i \rangle \sigma_{\mu\nu}+\cdots \, ,
\end{eqnarray}
\begin{eqnarray}
S_{ij}(x)&=& \frac{i\delta_{ij}\!\not\!{x}}{ 2\pi^2x^4}
-\frac{\delta_{ij}m_s}{4\pi^2x^2}-\frac{\delta_{ij}\langle
\bar{s}s\rangle}{12} +\frac{i\delta_{ij}\!\not\!{x}m_s
\langle\bar{s}s\rangle}{48}-\frac{\delta_{ij}x^2\langle \bar{s}g_s\sigma Gs\rangle}{192}+\frac{i\delta_{ij}x^2\!\not\!{x} m_s\langle \bar{s}g_s\sigma
 Gs\rangle }{1152}\nonumber\\
&& -\frac{ig_s G^{a}_{\alpha\beta}t^a_{ij}(\!\not\!{x}
\sigma^{\alpha\beta}+\sigma^{\alpha\beta} \!\not\!{x})}{32\pi^2x^2} -\frac{\delta_{ij}x^4\langle \bar{s}s \rangle\langle g_s^2 GG\rangle}{27648}-\frac{1}{8}\langle\bar{s}_j\sigma^{\mu\nu}s_i \rangle \sigma_{\mu\nu}  +\cdots \, ,
\end{eqnarray}
\begin{eqnarray}
C_{ij}(x)&=&\frac{i}{(2\pi)^4}\int d^4k e^{-ik \cdot x} \left\{
\frac{\delta_{ij}}{\!\not\!{k}-m_c}
-\frac{g_sG^n_{\alpha\beta}t^n_{ij}}{4}\frac{\sigma^{\alpha\beta}(\!\not\!{k}+m_c)+(\!\not\!{k}+m_c)
\sigma^{\alpha\beta}}{(k^2-m_c^2)^2}\right.\nonumber\\
&&\left. -\frac{g_s^2 (t^at^b)_{ij} G^a_{\alpha\beta}G^b_{\mu\nu}(f^{\alpha\beta\mu\nu}+f^{\alpha\mu\beta\nu}+f^{\alpha\mu\nu\beta}) }{4(k^2-m_c^2)^5}+\cdots\right\} \, ,\nonumber\\
f^{\alpha\beta\mu\nu}&=&(\!\not\!{k}+m_c)\gamma^\alpha(\!\not\!{k}+m_c)\gamma^\beta(\!\not\!{k}+m_c)\gamma^\mu(\!\not\!{k}+m_c)\gamma^\nu(\!\not\!{k}+m_c)\, ,
\end{eqnarray}
and  $t^n=\frac{\lambda^n}{2}$, the $\lambda^n$ is the Gell-Mann matrix
\cite{PRT85,Pascual-1984,WangHuang3900}.
We retain the terms $\langle\bar{q}_j\sigma_{\mu\nu}q_i \rangle$ and $\langle\bar{s}_j\sigma_{\mu\nu}s_i \rangle$ come from the Fierz re-arrangements   of the
$\langle q_i \bar{q}_j\rangle$ and $\langle s_i \bar{s}_j\rangle$  to  absorb the gluons  emitted from other quark lines to  extract the mixed condensate   $\langle\bar{q}g_s\sigma G q\rangle$ and $\langle\bar{s}g_s\sigma G s\rangle$, respectively \cite{WangHuang3900}.  Then we compute  all the integrals to obtain the correlation function $\Pi(p)$   at the quark-gluon  level, and finally obtain the QCD spectral densities through   dispersion relation,
\begin{eqnarray}\label{QCD-rho}
 \rho^1_{QCD}(s) &=&\frac{{\rm Im}\Pi^1(s)}{\pi}\, , \nonumber\\
\rho^0_{QCD}(s) &=&\frac{{\rm Im}\Pi^0(s)}{\pi}\, .
\end{eqnarray}
      There are two $c$-quark propagators and three light quark propagators  in Eq.\eqref{QCD-Pi}, if each $c$-quark line emits a gluon and each light quark line contributes  a quark-antiquark  pair, we obtain an operator $G_{\mu\nu}G_{\alpha\beta}\bar{u}u\bar{d}d\bar{s}s$ of dimension 13, so we should take into account the vacuum condensates at least up to dimension $13$ in a consistent way. Furthermore, we take into account the terms proportional to the $m_s$ to account for the flavor   $SU(3)$ mass-breaking effects.    The vacuum condensates
  $\langle\bar{q} q\rangle^2\langle\bar{q}g_s\sigma Gq\rangle $, $\langle\bar{q} q\rangle\langle\bar{s} s\rangle\langle\bar{q}g_s\sigma Gq\rangle $,  $\langle\bar{q} q\rangle^2\langle\bar{s}g_s\sigma Gs\rangle $, $\langle\bar{q} q\rangle \langle\bar{q}g_s\sigma Gq\rangle^2 $, $\langle\bar{s} s\rangle \langle\bar{q}g_s\sigma Gq\rangle^2 $, $\langle\bar{q} q\rangle \langle\bar{q}g_s\sigma Gq\rangle\langle\bar{s}g_s\sigma Gs\rangle  $,
  $\langle \bar{q}q\rangle^3\langle \frac{\alpha_s}{\pi}GG\rangle$, $\langle \bar{q}q\rangle^2\langle \bar{s}s\rangle\langle \frac{\alpha_s}{\pi}GG\rangle$ neglected in Refs. \cite{WangZG-EPJC-1509-12,WangZG-NPB-1512-32,WangZhang-APPB} are of dimension $11$ and $13$ respectively, and  are  associated with the inverse Borel parameters $\frac{1}{T^2}$, $\frac{1}{T^4}$, $\frac{1}{T^6}$ and $\frac{1}{T^8}$, which manifest themselves at  the small values of the Borel parameter $T^2$ and play an important role in determining the Borel windows.   We take the truncations $n\leq 13$ and $k\leq 1$ in a consistent way, just like in the new analysis in Ref.\cite{WZG-penta-IJMPA},
the quark-gluon operators of the orders $\mathcal{O}( \alpha_s^{k})$ with $k> 1$ and dimension $n>13$ are  discarded.

Now we  match the hadron side with the QCD side of the correlation function $\Pi(p)$, take the quark-hadron duality below the continuum threshold  $s_0$, and  obtain  the  QCD sum rules:
\begin{eqnarray}\label{QCDSR}
2M_{-}\lambda_{-}^2\exp\left( -\frac{M_{-}^2}{T^2}\right)&=& \int_{4m_c^2}^{s_0}ds \,\rho_{QCD}(s)\,\exp\left( -\frac{s}{T^2}\right)\,  ,
\end{eqnarray}
where $\rho_{QCD}(s)=\sqrt{s}\rho_{QCD}^1(s)+\rho_{QCD}^{0}(s)$, the explicit expressions of the QCD spectral densities are neglected for simplicity.

We differentiate     Eq.\eqref{QCDSR} with respect to  $\frac{1}{T^2}$, then eliminate the
 pole residue $\lambda_{-}$ and obtain the QCD sum rules for
 the mass of the hidden-charm  pentaquark state with the strangeness $S=-1$,
 \begin{eqnarray}
 M^2_{-} &=& \frac{-\int_{4m_c^2}^{s_0}ds \frac{d}{d(1/T^2)}\, \rho_{QCD}(s)\,\exp\left( -\frac{s}{T^2}\right)}{\int_{4m_c^2}^{s_0}ds \, \rho_{QCD}(s)\,\exp\left( -\frac{s}{T^2}\right)}\,  .
\end{eqnarray}

\section{Numerical results and discussions}
We take  the standard values of the  vacuum condensates
$\langle\bar{q}q \rangle=-(0.24\pm 0.01\, \rm{GeV})^3$,  $\langle\bar{s}s \rangle=(0.8\pm0.1)\langle\bar{q}q \rangle$,
 $\langle\bar{q}g_s\sigma G q \rangle=m_0^2\langle \bar{q}q \rangle$, $\langle\bar{s}g_s\sigma G s \rangle=m_0^2\langle \bar{s}s \rangle$,
$m_0^2=(0.8 \pm 0.1)\,\rm{GeV}^2$, $\langle \frac{\alpha_s
GG}{\pi}\rangle=0.012\pm0.004\,\rm{GeV}^4$    at the energy scale  $\mu=1\, \rm{GeV}$
\cite{SVZ79-1,SVZ79-2,PRT85,ColangeloReview}, and  take the $\overline{MS}$ masses $m_{c}(m_c)=(1.275\pm0.025)\,\rm{GeV}$
 and $m_s(\mu=2\,\rm{GeV})=(0.095\pm0.005)\,\rm{GeV}$
 from the Particle Data Group \cite{PDG}.
Furthermore,  we take into account
the energy-scale dependence of  the quark condensates, mixed quark condensates and $\overline{MS}$ masses according to  the renormalization group equation \cite{Narison-mix},
 \begin{eqnarray}
 \langle\bar{q}q \rangle(\mu)&=&\langle\bar{q}q\rangle({\rm 1 GeV})\left[\frac{\alpha_{s}({\rm 1 GeV})}{\alpha_{s}(\mu)}\right]^{\frac{12}{33-2n_f}}\, , \nonumber\\
 \langle\bar{s}s \rangle(\mu)&=&\langle\bar{s}s \rangle({\rm 1 GeV})\left[\frac{\alpha_{s}({\rm 1 GeV})}{\alpha_{s}(\mu)}\right]^{\frac{12}{33-2n_f}}\, , \nonumber\\
 \langle\bar{q}g_s \sigma Gq \rangle(\mu)&=&\langle\bar{q}g_s \sigma Gq \rangle({\rm 1 GeV})\left[\frac{\alpha_{s}({\rm 1 GeV})}{\alpha_{s}(\mu)}\right]^{\frac{2}{33-2n_f}}\, ,\nonumber\\
  \langle\bar{s}g_s \sigma Gs \rangle(\mu)&=&\langle\bar{s}g_s \sigma Gs \rangle({\rm 1 GeV})\left[\frac{\alpha_{s}({\rm 1 GeV})}{\alpha_{s}(\mu)}\right]^{\frac{2}{33-2n_f}}\, ,\nonumber\\
m_c(\mu)&=&m_c(m_c)\left[\frac{\alpha_{s}(\mu)}{\alpha_{s}(m_c)}\right]^{\frac{12}{33-2n_f}} \, ,\nonumber\\
m_s(\mu)&=&m_s({\rm 2GeV} )\left[\frac{\alpha_{s}(\mu)}{\alpha_{s}({\rm 2GeV})}\right]^{\frac{12}{33-2n_f}}\, ,\nonumber\\
\alpha_s(\mu)&=&\frac{1}{b_0t}\left[1-\frac{b_1}{b_0^2}\frac{\log t}{t} +\frac{b_1^2(\log^2{t}-\log{t}-1)+b_0b_2}{b_0^4t^2}\right]\, ,
\end{eqnarray}
  where $t=\log \frac{\mu^2}{\Lambda^2}$, $b_0=\frac{33-2n_f}{12\pi}$, $b_1=\frac{153-19n_f}{24\pi^2}$, $b_2=\frac{2857-\frac{5033}{9}n_f+\frac{325}{27}n_f^2}{128\pi^3}$,  $\Lambda=213\,\rm{MeV}$, $296\,\rm{MeV}$  and  $339\,\rm{MeV}$ for the flavors  $n_f=5$, $4$ and $3$, respectively  \cite{PDG}.
In the present work, we study  the hidden-charm pentaquark state $udsc\bar{c}$,  it is better to choose the flavor numbers $n_f=4$, and evolve those input parameters to a typical energy scale $\mu$, which satisfy the energy scale formula  $\mu=\sqrt{M_{P}-(2{\mathbb{M}}_c)^2}$ with the updated value ${\mathbb{M}}_c=1.82\,\rm{GeV}$ \cite{Wang1508-EPJC,WangHuang-EPJC-1508-12,WangZG-EPJC-1509-12,WangZG-NPB-1512-32,WZG-tetraquark-Mc,WangHuang3900,
Wang-tetra-formula-1,Wang-tetra-formula-2,Wang-tetra-IJMPA-1,Wang-tetra-IJMPA-2,Wang-tetra-IJMPA-3}. We tentatively  assign the $P_{cs}(4459)$ to be  the $[ud][sc]\bar{c}$ ($0$, $0$, $0$, $\frac{1}{2}$) pentaquark state and take into account the flavor $SU(3)$ mass-breaking effects, the optimal energy scale of the QCD spectral density is  $\mu=\sqrt{M_{P}-(2{\mathbb{M}}_c)^2}-m_s=2.4\,\rm{GeV}$.

In the QCD sum rules for  the  heavy, doubly-heavy, triply-heavy baryon states, hidden-charm pentaquark states and doubly-charm pentaquark states,  we usually choose the continuum threshold parameters as $\sqrt{s_0}=M_{gr}+ (0.5-0.8)\,\rm{GeV}$  \cite{Wang1508-EPJC,WangHuang-EPJC-1508-12,
 WangZG-EPJC-1509-12,WangZG-NPB-1512-32,WangZhang-APPB,WZG-penta-IJMPA,WangHbaryon-1,WangHbaryon-2,WangHbaryon-3,WangHbaryon-4,WangHbaryon-5,WangHbaryon-6,Wang-cc-baryon-penta},   where we use the subscript $gr$ to denote  the ground states.
 In the present work,  we choose the continuum threshold parameters to be  $\sqrt{s_0}= 5.15\pm0.10\,\rm{GeV}$, and verify the possible assignment of the $P_{cs}(4459)$ as the $[ud][sc]\bar{c}$ ($0$, $0$, $0$, $\frac{1}{2}$) pentaquark state.

\begin{figure}
\centering
\includegraphics[totalheight=7cm,width=10cm]{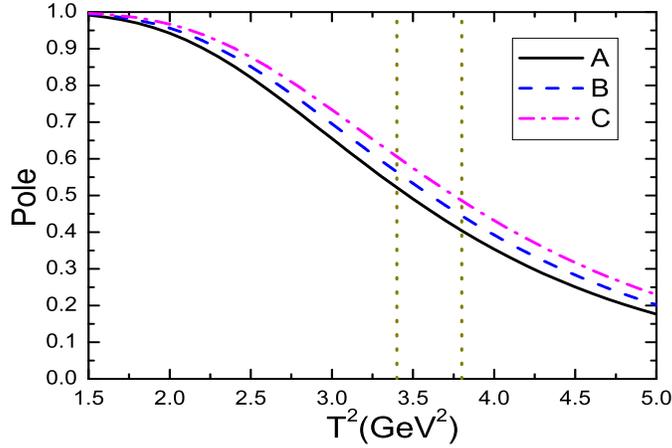}
  \caption{ The pole contributions  with variation of the  Borel parameter $T^2$, where the $A$, $B$ and $C$ correspond to the continuum threshold parameter $\sqrt{s_0}=5.05\,\rm{GeV}$, $5.15\,\rm{GeV}$ and $5.25\,\rm{GeV}$, respectively,   the region between the two vertical lines is the Borel window. }\label{Pole-Borel}
\end{figure}

 We obtain the  Borel  window $T^2=3.4-3.8\,\rm{GeV}^2$ via trial  and error, the pole contribution is about $(40-61)\%$, which is large enough to extract the pentaquark mass reliably, just as in our previous works on the hidden-charm pentaquark states and doubly-charm pentaquark states  \cite{Wang1508-EPJC,WangHuang-EPJC-1508-12,
 WangZG-EPJC-1509-12,WangZG-NPB-1512-32,WangZhang-APPB,WZG-penta-IJMPA,Wang-cc-baryon-penta}. In Fig.\ref{Pole-Borel}, we plot the pole contribution  with variation of the  Borel parameter $T^2$ at much larger range  than the Borel window.

\begin{figure}
\centering
\includegraphics[totalheight=7cm,width=10cm]{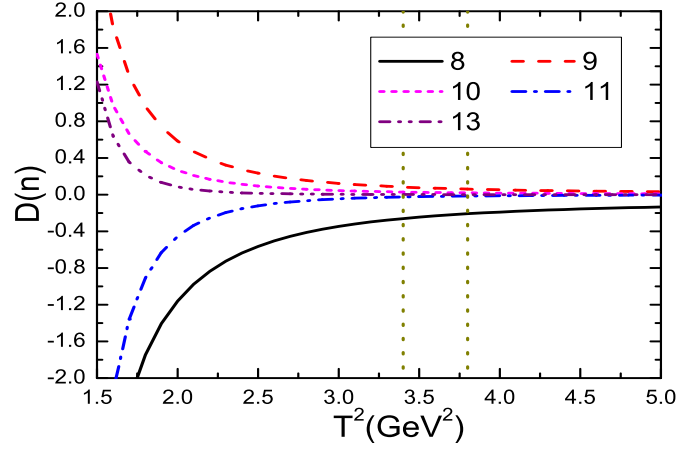}
  \caption{ The  contributions  of the higher dimensional  vacuum condensates $D(n)$ with variation of the  Borel parameter $T^2$, where the $n=8$, $9$, $10$, $11$ and $13$,   the region between the two vertical lines is the Borel window. }\label{OPE-Borel}
\end{figure}

In Fig.\ref{OPE-Borel}, we plot the contributions of the higher dimensional  vacuum condensates with variation of the  Borel parameter $T^2$. From the figure, we can see that the higher dimensional  vacuum condensates  manifest themselves at the region $T^2< 2.5\,\rm{GeV}^2$, we should choose the value $T^2> 2.5\,\rm{GeV}^2$. Their values decrease monotonously and quickly with the increase of the Borel parameter, in the Borel window $T^2=3.4-3.8\,\rm{GeV}^2$, the contributions of the higher dimensional vacuum condensates are  $D(8)=-(21-26)\%$, $D(9)=(6-8)\%$, $D(10)=(2-3)\%$, $D(11)=-(1-2)\%$, $D(13)\ll1\%$, the convergent behavior is very good.

\begin{figure}
\centering
\includegraphics[totalheight=7cm,width=10cm]{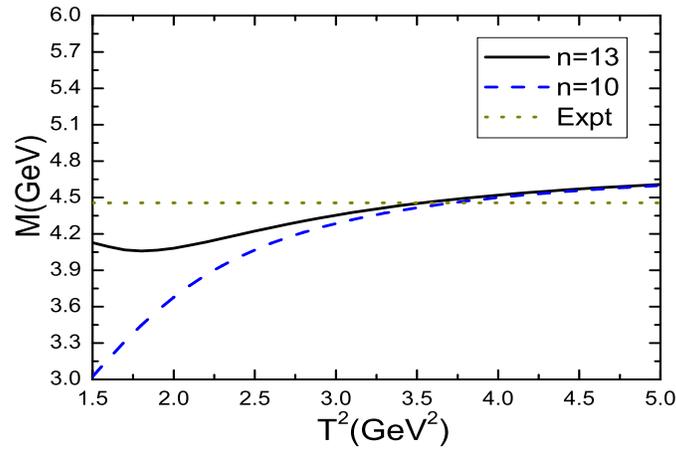}
  \caption{ The predicted  mass  with variation of the  Borel parameter $T^2$, where the $n=10$ and $13$ denote truncations of the operator product expansion, the expt denotes the experimental value. }\label{mass-D10-fig}
\end{figure}

The higher dimensional vacuum condensates play an important role in obtaining the flat platform, especially those associated with the inverse Borel parameters  $\frac{1}{T^2}$, $\frac{1}{T^4}$, $\frac{1}{T^6}$ and $\frac{1}{T^8}$. In Fig.\ref{mass-D10-fig}, we plot the predicted pentaquark mass  with variation of the  Borel parameter $T^2$ with the  truncations of the operator product expansion up to the vacuum condensates of dimensions $n=10$ and $13$, respectively.  From the figure, we can see that the vacuum condensates of dimensions $11$ and $13$ play an important role to obtain the flat platform, we should take them into account in a consistent way.

\begin{figure}
\centering
\includegraphics[totalheight=7cm,width=10cm]{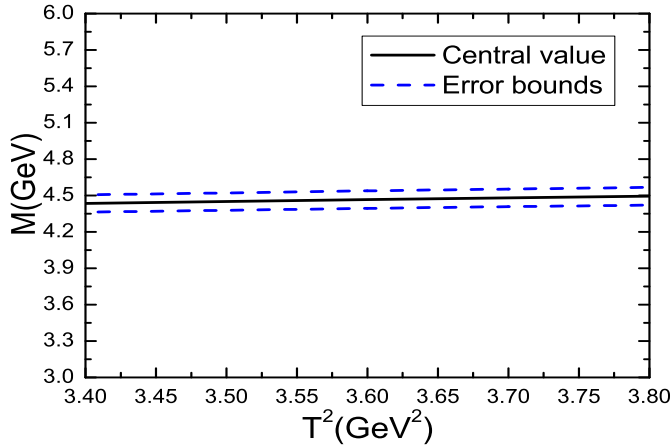}
  \caption{ The mass  with variation of the  Borel parameter $T^2$. }\label{mass}
\end{figure}

\begin{figure}
\centering
\includegraphics[totalheight=7cm,width=10cm]{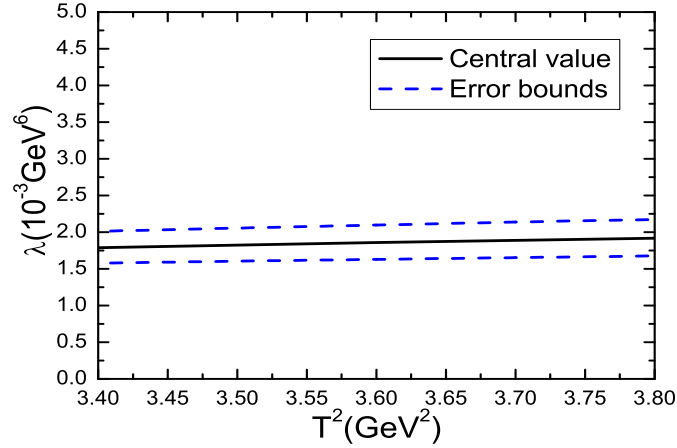}
  \caption{ The pole residue with variation of the  Borel parameter $T^2$. }\label{pole-residue}
\end{figure}

Now we take into account  all uncertainties  of the input   parameters,
and obtain  the mass and pole residue of
 the   hidden-charm pentaquark state $udsc\bar{c}$ with the negative-parity, which are shown explicitly in Figs.\ref{mass}-\ref{pole-residue},
\begin{eqnarray}
M_P&=&4.47\pm0.11\,\rm{GeV}\, ,\nonumber\\
\lambda_P&=&\left(1.86\pm0.31\right)\times 10^{-3}\,\rm{GeV^6}\, .
\end{eqnarray}
The predicted mass  $M_P=4.47\pm0.11\,\rm{GeV}$ is in excellent agreement with the experimental data $ 4458.8 \pm 2.9 {}^{+4.7}_{-1.1} \mbox{ MeV}$ from the  LHCb collaboration  \cite{LHCb-Pcs4459,LHCb-Pcs4459-2012}, and supports assigning the $P_{cs}(4459)$   as the $[ud][sc]\bar{c}$ ($0$, $0$, $0$, $\frac{1}{2}$) pentaquark state
with the spin-parity $J^P={\frac{1}{2}}^-$. In Ref.\cite{WZG-penta-IJMPA}, we observe that the $P_c(4312)$ can be assigned to be the $[ud][uc]\bar{c}$ ($0$, $0$, $0$, $\frac{1}{2}$) pentaquark state with the spin-parity $J^P={\frac{1}{2}}^-$. The flavor  $SU(3)$ mass-breaking effect is about $\Delta=m_s=147\,\rm{MeV}$, the estimations presented in Table \ref{mass-1508-et al-Pcs} are reasonable and reliable. In Refs.\cite{WangZG-EPJC-1509-12,WangZG-NPB-1512-32}, we study the $J^P={\frac{1}{2}}^\pm$ and ${\frac{3}{2}}^\pm$ hidden-charm pentaquark states with the strangeness $S=0$, $-1$, $-2$ and $-3$ in a systematic way by carrying out the operator product expansion up to the vacuum condensates of dimension $10$ and choosing the old value ${\mathbb{M}}_c=1.80\,\rm{GeV}$, and observe that the flavor  $SU(3)$ mass-breaking effects  are $\Delta=(90-130)\,\rm{MeV}$ for the negative-parity pentaquark states. The new analysis supports a larger flavor  $SU(3)$ mass-breaking effect.

The constituent quark models have given many successful descriptions of the hadron spectroscopy, the simple constituent quark mass plus hyperfine spin-spin interaction model works well for the ground state mesons ($M$) and baryons ($B$),
\begin{eqnarray}
M_M=\sum\limits_{i=1,2}\, m_i+\frac{C_{12}}{m_1 m_2}\,\vec{s}_1 \cdot \vec{s}_2\, ,\nonumber\\
M_B=\sum\limits_{i=1,2,3}\,m_i +\sum\limits_{i<j}\,\frac{C_{ij}}{m_i m_j}\,\vec{s}_i \cdot \vec{s}_j\, ,
\end{eqnarray}
where the $m_i$  are the constituent quark masses, the $\vec{s}_i$ are the quark spins,  the coefficients $C_{ij}$ can be fitted phenomenologically \cite{mass-spin-M}.  Accordingly, we can study the ground state hidden-charm tetraquark ($T$) and pentaquark ($P$) states  with the simple constituent diquark mass plus hyperfine spin-spin interaction model,
\begin{eqnarray}
M_T= m_{\mathcal{ D}}+m_{\mathcal{ \bar{D}}}+\sum\limits_{i<j}\,\widetilde{C}_{ij}\,\vec{s}_i \cdot \vec{s}_j\, ,\nonumber\\
M_P= m_{\mathcal{ D}}+m_{\mathcal{ D}^\prime}+m_{\bar{c}}+\sum\limits_{i<j}\,\widetilde{C}_{ij}\,\vec{s}_i \cdot \vec{s}_j\, ,
\end{eqnarray}
where the coefficients $\widetilde{C}_{ij}=\frac{C_{ij}}{m_im_j}$, the $m_D$  are the constituent diquark masses,  and the $\vec{s}_i$ are the quark spins \cite{Karliner-B-cs,Ali-EPJC-2018,Karliner-S-junction}.
 If we want to take into account the light flavor $SU(3)$-breaking effects satisfactorily, we have to introduce additional parameters, the bounding terms or string-junctions beyond the quark masses, diquark masses and  coefficients $\widetilde{C}_{ij}$  \cite{Karliner-B-cs,Karliner-S-junction}. The parameters $\widetilde{C}_{ij}$ in the mesons,  baryons, tetraquark states and pentaquark states are not necessary to be the same, as they correspond to different quark structures and wave functions, the $SU(3)$ breaking effects in the mesons,  baryons, tetraquark states and pentaquark states can be quite different. Generally speaking, they can be fitted phenomenologically when the experimental data accumulate. For the traditional charmed baryon states, the  flavor $SU(3)$ mass-breaking effects due to the $s$-quark are $\Delta\approx 190\,\rm{MeV}$ and $120\,\rm{MeV}$ for the flavor antitriplets and sextets respectively from the Particle Data Group \cite{PDG}, which can be accounted  for satisfactorily by adding the bonding terms \cite{Karliner-B-cs}. For the hidden-charm pentaquark states, there are only five candidates $P_c(4312)$, $P_c(4380)$, $P_c(4440)$, $P_c(4457)$ and $P_{cs}(4459)$, their quantum numbers such as the spin, parity, $\cdots$, have not been determined yet, it is difficult to estimate the $SU(3)$ breaking effects in a simple and intuitive way. In the present work, we study the $SU(3)$ breaking effects with the QCD sum rules, and account for the $SU(3)$ breaking effects by the $s$-quark mass $m_s$, $s$-quark condensate $\langle\bar{s}s\rangle$ and $s$-quark mixed condensate $\langle\bar{s}g_s\sigma Gs\rangle$ , and obtain the mass gap about $\Delta=147\,\rm{MeV}$, and make crude estimations of the mass-spectrum of the hidden-charm pentaquark states with the strangeness $S=-1$,   direct but tedious calculations are still needed to make robust predictions.

\section{Conclusion}
  In this article, we study the scalar-diquark-scalar-diquark-antiquark type $udsc\bar{c}$ pentaquark state with the QCD sum rules in details. We take into account  all the    vacuum condensates up to dimension $13$ in a consistent way with the truncations $\mathcal{O}(m_s^k)$ for $k\leq1$, and use the energy scale formula $\mu=\sqrt{M_{P}-(2{\mathbb{M}}_c)^2}-m_s$ with the updated effective $c$-quark mass ${\mathbb{M}}_c=1.82\,\rm{GeV}$ and subtract the mass-breaking effect due to the small $s$-quark mass  to choose  the optimal  energy scales of the QCD spectral densities. The predicted mass $M_P=4.47\pm0.11\,\rm{GeV}$ is in excellent agreement with the experimental data $ 4458.8 \pm 2.9 {}^{+4.7}_{-1.1} \mbox{ MeV}$ from the  LHCb collaboration, and supports assigning the $P_{cs}(4459)$   to be the $[ud][sc]\bar{c}$ ($0$, $0$, $0$, $\frac{1}{2}$) pentaquark state
with the spin-parity $J^P={\frac{1}{2}}^-$. We take into account the flavor   $SU(3)$ mass-breaking effect about $150\,\rm{MeV}$, and estimate the mass spectrum of the diquark-diquark-antiquark type $udsc\bar{c}$ pentaquark states with the strangeness $S=-1$.

\section*{Acknowledgements}
This  work is supported by National Natural Science Foundation, Grant Number  11775079.

\end{document}